%% file: main.tex
\documentclass[10pt, conference, a4paper]{IEEEtran}

\usepackage{cite}

\ifCLASSINFOpdf
  \usepackage[pdftex]{graphicx}
\else
  \usepackage[dvips]{graphicx}
\fi
\usepackage[tight,footnotesize]{subfigure}
\usepackage{url}

\usepackage[colorinlistoftodos, obeyDraft]{todonotes}

\usepackage{amsmath}

\begin{document}
%
\title{Radio Irregularity Model in OMNeT++}

\author{\IEEEauthorblockN{Behruz Khalilov, Anna F{\"o}rster, Asanga Udugama\vspace{0.2cm}}
\IEEEauthorblockA{Sustainable Communications Networks, 
University of Bremen,
Germany\\
Email: \{bk $|$ afoerster $|$ adu \}@comnets.uni-bremen.de}

}


%


\maketitle

\begin{abstract}

Radio irregularity is a non-negligible phenomenon that has an impact on protocol performances. For instance, irregularity in radio range leads to asymmetric links that cause the loss of packets in different directions. In order to investigate its effect, the Radio Irregularity Model (RIM) is proposed that takes into account the irregularity of a radio range and estimates path losses in an anisotropic environment. The purpose of this paper is to provide details of the RIM model developed in the INET Framework of the OMNeT++ simulator that can be used to investigate the impact of radio irregularity on protocol performance.  
   

\end{abstract}


%
\IEEEpeerreviewmaketitle

\input{intro.tex}

\input{impl.tex}
\input{param.tex}

\input{conclusion.tex}

\bibliographystyle{IEEEtran}
%

\end{document}

%% file: intro.tex

\section{Introduction}
\label{sec:intro}

\begin{figure}[htbp]
  \begin{center}
    \leavevmode
    \includegraphics[width=1.0\columnwidth]{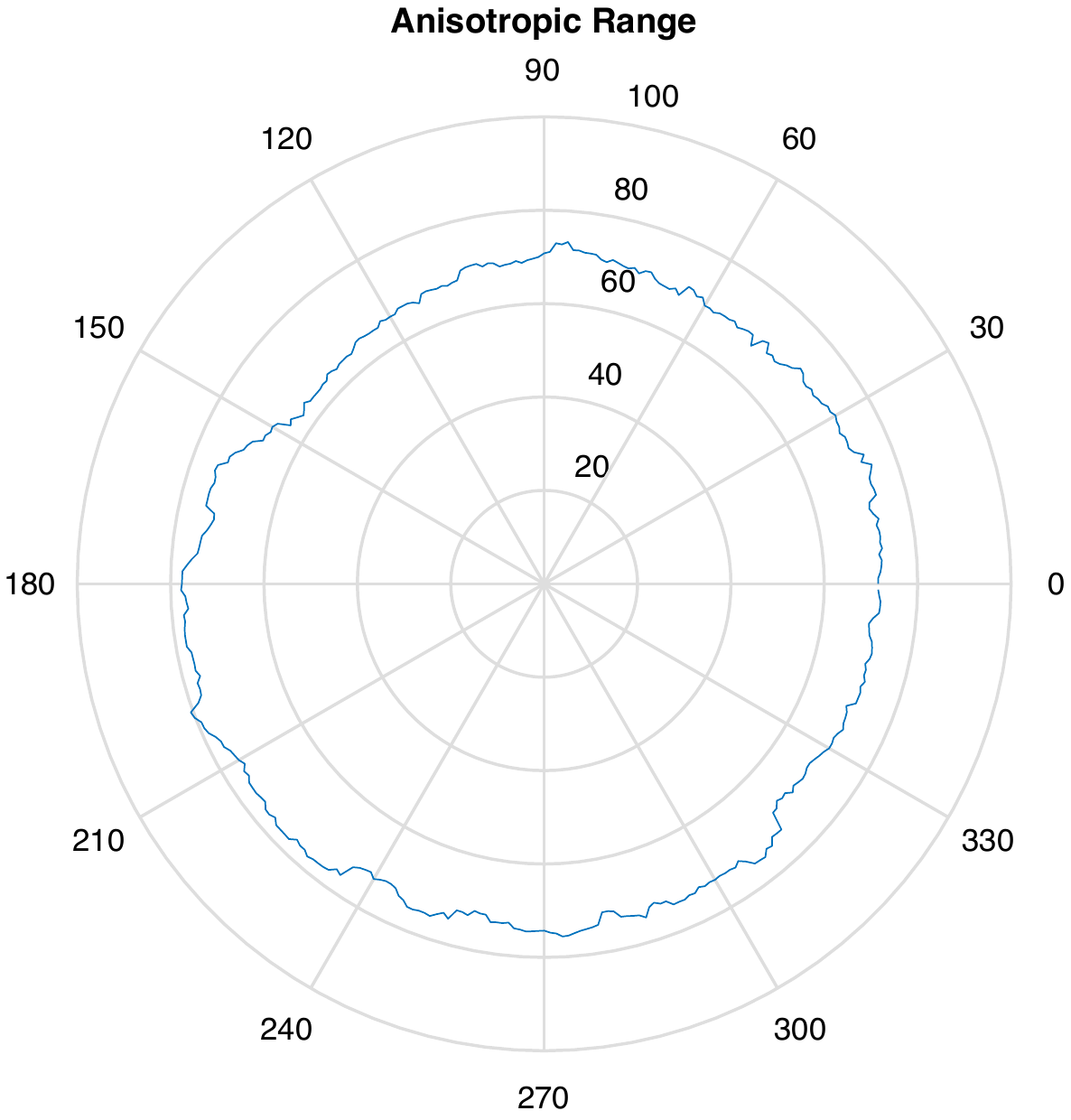}
    \caption{Path loss variations per unit degree change in the direction of radio propagation}
    \label{fig:plvariation}
  \end{center}
\end{figure}

Anisotropic properties of the propagation media and the heterogeneous properties of devices are two major sources of radio irregularity that leads to unstable communication between two or more devices. The impact of radio irregularity on protocol performances can be investigated either through a running system or via a simulation environment. The first approach is not feasible due to its complexity. Therefore, simulation environment is widely used for this purpose. However, when considering radio propagation, most of the existing simulation models assume a spherical radio pattern that leads to an inaccurate estimation. The authors of \cite{Zhou:2006} have proposed the Radio Irregularity Model (RIM) to cater for such irregularity that can be used for evaluation of an actual communication range of a sensor node dependent on medium and illustrate the variation of path loss in different directions. A key parameter of RIM is the Degree of Irregularity (DOI). DOI is used to describe the irregularity of the radio range and is defined as "\emph{the maximum path loss percentage variation per unit degree change in the direction of radio propagation}" \cite{Zhou:2006} (Figure~\ref{fig:plvariation}). A further aspect of RIM is the use of the Weibull distribution \cite{Weibull:1951} to model the variance of received signal strength of the different directions. The INET Framework of OMNeT++ provides a number of radio propagation models. The work presented here describes the RIM simulation model developed to supplement the other propagation models available in the INET Framework.

%% file: impl.tex
\section{INET Implementation of RIM}
\label{sec:imple}
The RIM functionality, i.e., \textbf{RIMFading}, is implemented in the INET Framework\footnote{INET Framework website: \url{https://inet.omnetpp.org/}} by extending the \textbf{FreeSpacePathLoss} model. It contains the following functionality.

\begin{enumerate}
  \item $K_{i}$, the coefficient of irregularity is generated once at the start of the simulation run for each node. \\
		The iteration always starts from a constant point and this point is north or origin point of a transmitter:
  \begin{equation*}
    K_{i}=\begin{cases}
      1; & \text{$i = 0$}.\\
      K_{i-1} \pm Rand * DOI ; & \text{$ 0 < i < 360^{0}$ }
      \end{cases}
  \end{equation*}
   \hspace{10pt}$|K_{0}-K_{359}| \leq DOI$
   \vspace{0.5em}
  \item Compute path loss using Free Space Path Loss (FSPL). Any other path loss or fading models can be used instead of the FSPL model.
  \item Calculate angle(s) between sender and receiver. By default RIMFading uses 2D model since the function that calculates the angle considers xy-plane and provides the value of the resulting angle for obtaining corresponding $K_{i}$ value. 
  \item Multiply FSPL value with the corresponding $K_{i}$ to obtain: $ DOIAdjusted Path Loss = Path Loss * K_{i} $
\end{enumerate}

Random numbers are generated using the Weibull distribution. Statistical analysis of empirical data collected from experiments \cite{Zhou:2006} have shown that the variation of received signal strength in different direction fits the Weibull distribution. 

\begin{figure}[htbp]
  \begin{center}
    \leavevmode
    \includegraphics[width=1.0\columnwidth]{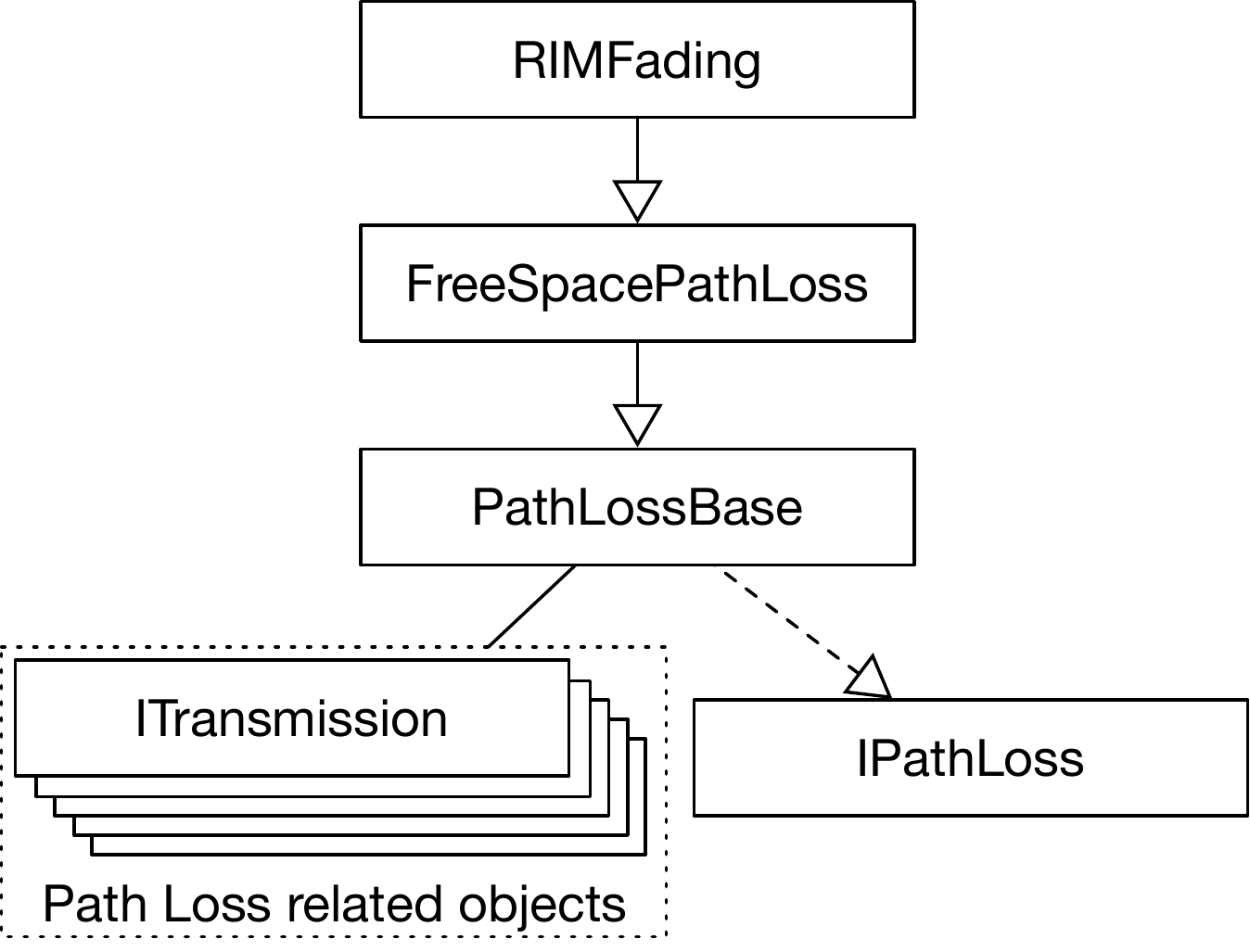}
    \caption{Class inheritance diagram of the RIM implementation}
    \label{fig:inhdiagram}
  \end{center}
\end{figure}

Figure~\ref{fig:inhdiagram} shows the class inheritance diagram of the RIM implementation. It extends the functionality of the FSPL model to include the RIMFading.

%% file: param.tex

\section{Parameters}

The RIMFading model has a number of configurable parameters that are defined in the NED file:

\begin{itemize}
\item \textbf{a}: The scale parameter of the Weibull distribution (default is 1.5).
\item \textbf{b}: The shape parameter of the Weibull distribution (default is 1).
\item \textbf{DOI}: The degree of irregularity (default is 0.006). 
\end{itemize}

\vfill
\pagebreak

The default values for the shape (\textbf{a}) and scale (\textbf{b}) parameters of the Weibull distribution are selected to generate the required range of random numbers that are used to compute the coefficient of irregularity. The default value of \textbf{DOI} is based on the recommendation given in \cite{Zhou:2006}.

%
%
%
%
%

%% file: conclusion.tex

\section{Conclusion}
\label{sec:conclusion}

The work presented in this paper relates to the development of the RIM simulation model for the INET Framework to simulate the variance of path loss with respect to the direction of propagations. The model developed in this work is released at Github (\textbf{https://github.com/ComNets-Bremen/RIMFading}) under a GPL License.